\documentclass{article}
\usepackage{spconf,amsmath,graphicx}
\usepackage{mntzpreamble}
\usepackage{scrextend}
\usepackage{makecell}
\usepackage{textpos}

\title{COMPRESSIVE REGULARIZED DISCRIMINANT ANALYSIS OF HIGH-DIMENSIONAL DATA WITH APPLICATIONS TO MICROARRAY STUDIES}

\name{Muhammad Naveed Tabassum and Esa Ollila \thanks{The research was partially supported by the Academy of Finland grant no. 298118 which is gratefully acknowledged.}}
\address{Aalto University, Dept. of Signal Processing and Acoustics, P.O. Box 15400, FI-00076 Aalto, Finland}

\begin{document}
%
\maketitle
\begin{textblock*}{\textwidth}(0.15\textwidth,-7cm)
\begin{small}
To appear in Proceedings of the IEEE International Conference on Acoustics, Speech and Signal Processing, 2018.\end{small}
\end{textblock*}
\vspace{-5ex}

\begin{abstract}
We propose a modification of linear discriminant analysis, referred to as {\sl compressive regularized  discriminant analysis (CRDA)}, for analysis of high-dimensional datasets. CRDA is especially designed for feature elimination purpose and can be used as gene selection method in microarray studies.   CRDA lends ideas from $\ell_{q,1}$ norm minimization algorithms in the multiple measurement vectors (MMV) model and utilizes joint-sparsity promoting hard thresholding for feature elimination. 
A regularization of the sample  covariance  matrix is also needed as we consider the challenging scenario  where the number of features (variables) is comparable or exceeding the sample size of the training dataset.
A simulation study and four examples of real life microarray datasets evaluate the performances of CRDA based classifiers. 
Overall, the proposed method gives fewer misclassification errors than its competitors, while at the same time achieving accurate feature selection. 
\end{abstract}


\section{Introduction} \label{sec:intro}

Sparse signal approximations are widely used in many applications such as regression or classification where variable-selection (i.e., ranking and selection of features) aims at reducing the number of variables (or features) without sacrificing accuracy measured by the test error.  Reduction in the set of features facilitates  interpretation as well as  stabilizes estimation.  
This is often deemed necessary in the high-dimensional (HD) context where the number of features, $\pdim$, is often several magnitudes larger than the number of observations, $\ndim$, in the training dataset (i.e., $\pdim \gg \ndim$).

Many classification techniques assign a $\pdim$-dimensional observation $\x$ to one of the $G$ classes (groups or populations) based on the following rule 
\beq\label{eq:ldarule} 
\x  \in \texttt{group}\, \big[\, \gt = \arg \max_{g} \: d_{g}(\x) \,\big],
\eeq
where $d_g(\x)$ is called the {\sl discriminant function} for population $g \in \{1,\ldots, G\}$. In  linear discriminant analysis (LDA),  $d_g(\x)$ is a linear function of $\x$,  $d_g(\x)= \x^\top \beb_g + c_g$, for some constant $c_g \in \R$ and vector $\beb_g \in \R^\pdim$. The vector $\beb_g=\beb_g(\Sig)$ depends on the unknown covariance matrix $\Sig$ of the populations (via its inverse matrix) which is commonly estimated by the pooled sample covariance matrix (SCM). In the HD setting, the SCM is no-longer invertible, and therefore regularized SCM (RSCM) $\hat \Sig$ is used  for constructing an estimated discriminant function $\hat d_g(\x)$.  Such approaches are commonly referred to as regularized LDA methods, which we refer shortly as RDA. See e.g.,  \cite{guo2006regularized,tibshirani2003NSC,witten2011plda, sharma2014feature, neto2016regularized}.

Next note that if the $i$-th entry of $\beb_g$ is zero, then  the $i$-th feature does not contribute in the classification of $g$-th population.
To eliminate unnecessary features, many authors have proposed to shrink $\beb_g$ using element-wise soft-thresholding, e.g., as in  shrunken centroids (SC)RDA method \cite{guo2006regularized}.
These methods are often difficult to tune because the shrinkage threshold parameter is the same across all groups, but different populations would often benefit from different shrinkage intensity.
Consequently, they tend to yield rather higher false-positive (FP) rates.  
 
Element-wise shrinkage does not achieve {\it simultaneous} feature selection as the eliminated feature from group $i$ may still affect the discriminant function of group $j$. 
In this paper,  we propose  {\sl compressive regularized  discriminant analysis (CRDA)} that promotes simultaneous {\it joint-sparsity} to pick fewer and differentially expressed variables.  CRDA lends ideas from mixed $\ell_{q,1}$ norm minimization in the multiple measurement vectors (MMV) model \cite{duarte2011structured}, 
which is an extension of compressed sensing model to multivariate case.   
CRDA uses  $\ell_{q,1}$-norm based hard-thresholding  which  has the advantage of having a shrinkage parameter that is much easier to tune: namely, joint-sparsity level $K \in \{ 1, \ldots, \pdim\}$ instead of shrinkage threshold $\Delta \in [0, \infty)$ as in SCRDA. Our approach also employs a different RSCM estimator compared to SCRDA.  The used RSCM has the benefit of  being able to attain the minimum mean squared error \cite{ledoit2004well, ollila2017optimal} for an appropriate choice of the regularization parameter.   The optimal pair of the tuning  parameters can be found via cross validation (CV), but we also propose a computationally simpler approach that uses the RSCM  proposed in \cite{ollila2017optimal}.  This facilitates the computations considerably as only a single variable, the joint-sparsity level $K$, needs to be tuned.

The paper is organized as follows. Section~\ref{sec:RLDA} describes RDA and SVD based inversion of the used RSCM. In Section~\ref{sec:CRDA}, the proposed CRDA as well as our tuning parameter selection criteria is introduced.  Section~\ref{sec:results} provides the results on simulation studies which explore both the feature-selection capability and misclassification errors of CRDA, and the competing methods.  Classification results on  
four real microarray datasets  are also provided.  
Section~\ref{sec:concl} concludes the paper.

\section{Regularized LDA} \label{sec:RLDA}

We are given a $\pdim$-variate random vector $\x$ which we need to classify into one of the $G$ classes or populations. 
In LDA, one assumes that the class populations are $\pdim$-variate multivariate normal (MVN) with a common positive definite symmetric covariance matrix $\Sig$ over each class but distinct class mean vectors $\m_g \in\mathbb{R}^{\pdim}$, $g=1,\ldots, G$.  The problem is then to classify  $\x$ to one of the MVN populations, $\mathcal N_p( \m_g, \Sig)$, $g=1,\ldots, G$.  Sometimes prior knowledge is available on proportions of each population and we denote by $p_g$, $g=1,\ldots,G$, the  prior probabilities of the classes ($\sum_{g=1}^G p_g = 1$). 
LDA uses the rule \eqref{eq:ldarule} with discriminant function
\[
d_g(\x)=   \x^\top \beb_g - \frac 1 2    \m_g ^\top  \beb_g   + \ln p_g,
\]
where $\beb_g = \Sig^{-1} \m_g$ for $g=1,\ldots, G$.  

The LDA rule involves a set of unknown parameters, the class mean vectors $\m_g$ and the covariance matrix $\Sig$. 
These are estimated from the {\sl training dataset}  $\X=(\x_1 \, \cdots \, \x_n)$ that consists of $n_g$ observations from each of the classes ($g=1,\ldots, G$).  
Let $c(i)$ denote the class label associated with the  $i$-th observation, so $c(i) \in \{ 1, \ldots, G\}$.  Then $ n_g = \sum_{i=1}^n \mathsf{I}(c(i)=g)$ is the number of observations belonging to $g$-th population,  and we denote by  $\pi_g=n_g/n$  the relative sample proportions. 
We assume observations in the training dataset are centered by the sample mean vectors of the classes, 
\vspace{-0.2ex}
\beq \label{eq:muh}
\muh_g  = \overline{\x}_g = \frac{1}{n_g}\,\sum_{ c(i)=g} \x_i.  
\eeq 
 Since $\X$ is centered,  the pooled (over groups) sample covariance matrix (SCM) can be written simply as 
 \[
 \S =  \frac{1}{n} \,\Xs \Xs^\top.
 \] 
In practice, an observation $\x$ is classified using an estimated discriminant function, 
\beq\label{eq:edf} 
\hat d_g(\x) =  \x^\top \bebh_g - \frac 1 2  \muh_g^\top \bebh_g     + \ln \pr_g, 
\eeq
where $\bebh_g = \Sigh^{-1} \muh_g$, $g=1,\ldots, G$ and $\Sigh$ is an estimator of  $\Sig$. 
Note that in \eqref{eq:edf}  the prior probabilities $p_g$-s are  replaced by their estimates, $\pr_g$-s.  
Commonly, the pooled SCM $\bo S$ is used as an estimator $\Sigh$. 
Since we are in the regime, where $p \gg n$, the pooled SCM is no longer invertible and hence can not be used in \eqref{eq:edf}. 
To avoid the singularity of the estimated covariance matrix, a commonly used approach in the literature (cf. \cite{ledoit2004well,ollila2017optimal}) is to 
use  a {\paino regularized SCM (RSCM)}, 
\beq \label{eq:rscm} 
\Sigt =  \al \,\S + (1- \al) \, \eta \bo I 
\eeq 
where $\eta = \tr(\S)/\pdim$. 
SCRDA \cite{guo2006regularized}  uses  an estimator  $\Sigt =  \al \S + (1- \al) \bo I$. However,  \eqref{eq:rscm} has some theoretical justification  since with an appropriate (data dependent) choice $\hat \al$, the obtained RSCM in  \eqref{eq:rscm} will be a consistent minimum mean squared error (MMSE) estimator of $\Sig$. Such choices of $\al$  have been proposed, e.g., in \cite{ledoit2004well} and in \cite{ollila2017optimal}.

In the HD setup, the main computational burden is related with inverting the matrix $\Sigt$ in \eqref{eq:rscm}. 
The inversion can be done using the SVD-trick, as follows \cite{guo2006regularized, friedman2001elements}. The SVD of $\X$ is  
\[
\Xs= \Us \Ds \Vs^\top,
\]
where $\Us  \in \R^{p \times m}$, $\Ds \in \R^{m \times m}$, $\Vs \in \R^{n \times m}$ and $m = \mathrm{rank}(\Xs)$. 
Direct computation of SVD is time consuming and the trick is that $\V$ and $\D$ can be computed first from SVD of $\Xs^\top \Xs = \tilde \V \tilde \D^2 \tilde \V^\top$,  which is only an $n \times n$ matrix. Here  $\tilde \V$ is an orthogonal $ \ndim \times \ndim$  matrix whose first $m$-columns are $\Vs$ and $\tilde \D$ is an $ \ndim \times \ndim$ diagonal matrix whose upper left corner $m \times m$ matrix is $\Ds$. 
After we compute  $\Vs$ and $\Ds$ from SVD of $\Xs^\top \Xs$, we may compute $\Us$ from $\X$ by $\Us = \X \Vs \Ds^{-1}$. 
Then, using the SVD representation of the SCM, $\S = (1/n) \Ub\D^2 \Ub^\top$,  and simple algebra, one obtains a simple formula for the inverse:
\begin{align}\label{eq:RSCMinv} 
\Sigit &= \Us \bigg[ \Big( \frac{\al}{n}\,\Ds^2 + (1-\al)  \eta \bo I_m  \Big)^{-1} - \frac{1}{(1-\al)\eta} \bo I_m \bigg] \,\Us^\top \notag \\  & \qquad \quad+ \frac{1}{(1-\al) \eta} \bo I_\pdim,
\end{align}
where $\eta = \tr(\S)/\pdim = \tr(\Ds^2)/n\pdim$.
This reduces the complexity from  $\mathcal{O}(\pdim^3)$ to $\mathcal{O}(\pdim \ndim^2)$ which is a significant saving in $p \gg n$ case.

\section{Compressive RDA } \label{sec:CRDA}

\subsection{Proposed CRDA Approach} 

In order to explain the proposed  compressive RDA approach,  
we first write the discriminant rule in vector form  as 
\begin{align} 
\bo d(\x) &= (d_1(\x), \ldots, d_G(\x) )  \notag  \\
&=  \x^\top \B  - \frac 1 2   \, \mathrm{diag} \big(  \M^\top  \B  \big)    + \ln \bo p, \label{eq:koe}
\end{align} 
where $\ln \bo p = (\ln p_1,\ldots, \ln p_G)$, $\M  = \bmat \m_1 & \ldots & \m_G \emat$ and  $\B = ( \beb_1 \, \cdots \, \beb_G) = \Sig^{-1} \M$. 
Above notation $\mathrm{diag}(\cdot )$ extract the diagonal of the $G \times G$ matrix $\bo A$ into a vector, i.e., $\mathrm{diag}(\bo A)=(a_{11}, \ldots,a_{GG})$.
The discriminant function in \eqref{eq:koe} is linear in $\x$ with coefficient matrix $\B \in \R^{\pdim \times G}$. This means that if the $i$-th row  of the coefficient matrix  $\B$ is a zero vector $\bo 0$, then it implies that $i$-th predictor does not contribute to the classification rule and hence can be eliminated. If the coefficient matrix $\B$ is row-sparse, then the method can be potentially used as a simultaneous feature elimination procedure. 
In microarray data analysis, this means that gene $i$ does not contribute in the classification procedure and thus the row-sparsity of the coefficient matrix allows, at the same time,  identify differentially expressed genes. 

In the MMV model \cite{duarte2011structured}, the goal is to achieve simultaneous sparse reconstruction (SSR) of the signal matrix. The task is  to estimate the {\paino $K$-rowsparse} signal matrix $\B$, given an observed measurement matrix $\bo Y$ and  an (over complete) basis matrix (or dictionary) $\bom \Phi$. $K$-rowsparsity of $\B$ means that only $K$ rows of $\B$ contain non-zero entries. Commonly, this goal is achieved by $\ell_{q,1}$ mixed matrix norm minimization,  where 
\[
\| \B \|_{q,1}  =  \sum_{i=1}^\pdim \| \beb_{[i]} \|_q   , 
\] 
for some $q \geq 1$, where $\beb_{[i]}$ denotes the $i$-th row of $\B$. Values $q=1,2, \infty$ have been advocated in the literature. 
Many  SSR algorithms use {\paino hard-thresholding operator} $H_K(\cdot, q)$, defined as 
transform $H_\K(\B, q)$, which   retains the elements of the $\K$ rows of $\B$ that possess largest $\ell_{q}$ norm and set elements of the other rows to zero.  
This leads us to define our {\paino compressive RDA} discriminant function as 
\begin{align} 
\hat{\bo d}(\x) &= \big(\hat d_1(\x), \ldots, \hat d_G(\x) \big)  \notag  \\ &=  \x^\top  \hat \B  - \frac 1 2   \, \mathrm{diag} \big(  \hat \M^\top  \hat \B  \big)    + \ln \bom \pr, \label{eq:koe2}
\end{align} 
where $\ln \bom \pr = (\ln \pr_1,\ldots, \ln \pr_G)$, $\hat \M  = \bmat \muh_1 & \ldots & \muh_G \emat$ and  
\[
\hat \B =  H_K( \hat \Sig^{-1} \hat \M,q)
\] 
where $\Sigh$ has been defined in \eqref{eq:rscm} and $\muh_g$ are the sample mean vectors of the classes in \eqref{eq:muh}. 
Fast formula to compute $\hat \Sig^{-1}$ is given in \eqref{eq:RSCMinv}.

Next let us draw attention to SCRDA \cite{guo2006regularized} which uses $\Sigt =  \al \S + (1- \al) \bo I$ instead of  estimator in \eqref{eq:rscm}.  Another difference is in its use of element-wise soft-shrinkage.  Namely, SCRDA can also be written in the multivariate form  \eqref{eq:koe2}, but using 
\beq \label{eq:SCRDA} 
\hat \B =  \mathcal S_{\Delta}( \hat \Sig^{-1} \hat \M)
\eeq 
where $\mathcal S_\Delta(\cdot)$ is the soft-thresholding function that is applied element-wise to its matrix-valued argument. That is, the $(i,j)$-th element  $ \hat b_{ij}$ 
of $\hat \B $ in   \eqref{eq:SCRDA} is 
\[
\hat b_{ij} = \mathcal S_\Delta(  t_{ij})=  \mathrm{sign}(t_{ij})(|t_{ij}| - \Delta)_+
\]
where  $(t)_+=\max(t,0)$  for $t \in \R$   and $t_{ij}$ denotes the  $(i,j)$-th element of $\bo T= \hat \Sig^{-1} \hat \M$. 
One disadvantage of SCRDA is  the shrinkage thresholding parameter  $\Delta \in [0, \infty)$ which is  the same across all groups, and  different populations would often benefit from different shrinkage intensity. A sensible upper bound  of $\Delta$ is difficult to determine and is highly data dependent. 
The proposed CRDA on the other hand uses simple to tune joint-sparsity level $K \in \{1, 2, \ldots ,\pdim\}$  and has the benefit of offering simultaneous joint-sparse recovery, i.e., features are eliminated across all groups instead of group-wise.

\begin{table}[!t]
    \setlength\extrarowheight{2pt}
	\setlength{\textfloatsep}{-0.1cm}
	\centering
	\caption{Classification results for  the simulation setups I --III. Figures in bold-face indicate the best results  in each column. 
	For  setup III,  the false positive (FP) and detection rate (DR) are also reported. Results in paranthesis are obtained using  $(\alh_{ell}, \hat \K_\cv)$ instead of $(\alh_\cv, \hat \K_\cv)$.} \label{table:t1}
	\resizebox{0.49\textwidth}{!}
	{\begin{tabular}{ l | c c | c c }
			\cline{1-5}
			\multirow{2}{*}{\makecell{Methods}}  &\multicolumn{2}{c|}{Setup I}   &\multicolumn{2}{c}{Setup II} \\ 

			\cline{2-5}
			&TE &NFS    &TE &NFS \\
			\hline
			
			$\text{CRDA}^{\ell_1}$  &120 (116)  &165 (163)      &180 (174)  &105 (101) \\
			$\text{CRDA}^{\ell_2}$  &95 (94)    &126 (120)      &184 (182)  &96 (105)  \\
			$\text{CRDA}^{\ell_\infty}$  &\textbf{84 (81)}  &\textbf{112 (114)}      &185 (177)  &\textbf{94 (96)} \\
			
            PLDA    &117 &301    &\textbf{151}      &148  \\
            SCRDA   &97  &227    &291               &349  \\
            NSC     &89  &290    &277               &440  \\
			\hline

		& \multicolumn{4}{c}{ Setup III} \\
			\hline
			Methods &TE &NFS    &DR &FP \\
			\hline
			$\text{CRDA}^{\ell_1}$  &\textbf{46} (50)  &\textbf{205} (259)      &90 (\textbf{94})  &\textbf{12} (27) \\
			$\text{CRDA}^{\ell_2}$  &49 (\textbf{46})    &240 (\textbf{203})      &\textbf{92} (92)  &23 (\textbf{10})  \\
			$\text{CRDA}^{\ell_\infty}$  &50 (52)  &238 (252)      &89 (92)  &27 (27) \\
			SCRDA  &108  &282      &69  &51 \\
			\hline
\end{tabular} }
\end{table}

\begin{table*}[!t]
	\setlength\extrarowheight{2pt}
	\begin{minipage}{\textwidth}
    \centering
	\caption{Classification results for the four microarray datasets using 5-fold CV. Note that figures in bold-face indicate the best results. Results in parenthesis are obtained using  $(\alh_{ell}, \hat \K_\cv)$ instead of $(\alh_\cv, \hat \K_\cv)$. }  \label{table:t3rd}
	{\begin{tabular}{ l | c c | c c | c c | c c }
			\cline{1-9}
			\multirow{2}{*}{\makecell{Methods}}  &\multicolumn{2}{c|}{Ramaswamy {\it et al.} dataset}   &\multicolumn{2}{c|}{Yeoh {\it et al.} dataset}   &\multicolumn{2}{c|}{Sun {\it et al.} dataset}   &\multicolumn{2}{c}{Nakayama {\it et al.} dataset} \\ 

			\cline{2-9}
			&TE / 47 &NFS    &TE / 62 &NFS    &TE / 45 &NFS    &TE / 26 &NFS \\
			\hline
			$\text{CRDA}^{\ell_1}$  &10.6 (\textbf{9.9})  &2634 (4899)      &9.6 (7.5)  &2525 (4697) &12.5 (\textbf{12.9})  &23320 (27416) &8.3 (7.9) &2941 (6952) \\
			$\text{CRDA}^{\ell_2}$  &10.4 (10.3)    &2683 (3968)      &9.7 (\textbf{6.0}) &2273 (\textbf{4659})  &12.9 (13.3) &\textbf{20589} (23484) &7.9 (7.6) &3142 (7755) \\
			$\text{CRDA}^{\ell_\infty}$  &\textbf{10.3} (10.3)  &3405 (4530)      &\textbf{9.3} (6.5)  &\textbf{846} (4697) &\textbf{12.4} (13.5)   &21354 (\textbf{20207}) &7.6 (7.6) &\textbf{2719 (2340)} \\
			
            PLDA    &18.8 &5023     &NA &NA      &15.2 &21635 &4.4 &10479  \\
            SCRDA   &24   &14874    &NA &NA              &15.7 &54183 &\textbf{2.8} &22283   \\
            NSC     &16.3  &\textbf{2337}   &NA &NA    &15 &30005 &4.2 &5908   \\
			\hline
\end{tabular} }
\end{minipage}
\end{table*}

\subsection{Model (Parameters) Selection} \label{sec:params}

We employ $Q$-fold CV to estimate the optimal pair $(\alh_\cv,\hat \K_\cv)$ using a 2D grid of candidate values $\{\al_{i}\}_{i=1}^I \times \{\K_j\}_{j=1}^J$ of the tuning parameters, 
where $\al \in [0,1)$ and $\K \in  [1,p]=\{1,2, \ldots,\pdim\} \subset \mathbb{N}$. Often there are several pairs that yield the minimal cross-validation error from the training dataset and each pair can exhibit varying degree of sparsity (number of features selected).  Among them, we would prefer the pair that had minimal number of features. Since a pair with minimal CV error may not yield a classifier that is at the same time sparse, one may wish to set a lower bound for the number of features selected (NFS) in order to enhance the interpretability of the discriminant function. 

Let  $\varepsilon_\cv(\al, K) $ denote a CV error for a pair $(\al,K)$. 
To have a trade-off between a minimum (CV-based) training error $\varepsilon_\cv  \in [1,\ndim]$ and NFS, we use a threshold 
$\varepsilon^\thr = \max (0.15 \cdot n, \,\varepsilon_{\cv})$  and choose only the pairs which  have  CV error smaller than $\varepsilon^\thr$, i.e., 
pairs which verify  $\varepsilon_\cv(\al, K)  \leq \varepsilon^\thr$. From these pairs, the final optimal pair $(\alh_\cv,\hat \K_\cv)$ is chosen as the one that has the smallest NFS value. For finding the optimal pair, we utilize a uniform grid of  100 $K$-values and a uniform grid of 25 $\al$-values. 

 We compare the CV approach to computationally much lighter approach which uses the estimated  parameter $\alh_{ell}$ given in \cite{ollila2017optimal}. We note that value of  $\alh_{ell}$  can be  computed efficiently using the SVD trick. Given the optimal RSCM based on  $\alh_{ell}$ we then estimate the sparsity level $K$ using CV estimate  $\hat \K_{\cv}$. This reduces the computational cost significantly.

\section{Numerical examples} \label{sec:results}

The simulation study investigates the performance of CRDA based classifiers using different simulation setups commonly used in the RDA literature (e.g. in \cite{guo2006regularized, witten2011plda, ramey2013comparison, zhou2017gdrda}) and draw a comparison with the available results, against the nearest shrunken centroids (NSC) \cite{tibshirani2003NSC}, SCRDA \cite{guo2006regularized} and PLDA \cite{witten2011plda}.
For simulation setups I and II, we generate 1200 observations from MVN distribution, $\mathcal{N}_p (\m_g, \I_p)$, with equal probabilities for each of $G=4$ groups. The observations are divided into three sets: (i) the validation set with 100 observations finds the tuning parameters, (ii) then 100 observations in the training set estimate $\hat \Sig^{-1}$  and (iii) the rest 1000 form the test set for calculating misclassification test errors (TE). A total of $T= 100$ out of $\pdim=500$ features differ between the groups. In setup I, $\m_g$ contains $t=25$  nonzeros for each group $g$ and rest all zeros, i.e.,  $[\m_g]_i = 0.7$ for $t(g-1)+1\leq i \leq t(g-1)+t$ . While, $[\m_g]_i = \tfrac{g-1}{3}$ if $i \leq 100$ and zero otherwise for setup II. Table \ref{table:t1} lists the average of the TE and NFS for each classifier using 25 MC trials and 5-fold CV.

The third simulation setup resembles real  gene expression data. We generate $n=200$ training and $1000$ test observations each having $p=10,000$ features. All groups have equal probabilities and follow MVN distribution $\mathcal{N}_p (\m_g, \Sig_g)$ for $g=1,\ldots,G=3$.  We have  $\m_1 = \0_p$ and  
 $\m_2$ contains all zeros except first $200$-entries (i.e., true positives) with  value $1/2$ and $\m_3=-\m_2$. 
Each group employs following  block-diagonal auto-regressive covariance-structure 
\[
\ \Sig_{g} = \Sig^{(\rho_g)} \oplus \Sig^{(-\rho_g)} \oplus \cdots \oplus \Sig^{(\rho_g)} \oplus \Sig^{(-\rho_g)},
\]
where $\oplus$ indicates the direct sum (not the Kronecker sum) of $100$ block matrices having the AR$(1)$ covariance structure
\[
[\Sig^{(\rho_g)}]_{1\leq i, j\leq 100} =  \rho^{|i-j|}_g 
\]
where  $\rho_g$ is the correlation which is different for each group, namely,  $\rho_1= 0.5$, $\rho_2 = 0.7$ and $\rho_3 =0.9$. 
This setup mimics real microarray data as genes are correlated within a pathway and independent between the pathways. Table \ref{table:t1} reveals the higher accuracy of the proposed CRDA methods compared to SCRDA when measured by TE, NFS, detection rate (DR) and FP rates. The results are averaged over 10 Monte-Carlo trials using 10-fold CV. 

Next we do a comparison based on real microarray datasets. 
A summary of the used datasets is given below: 
\begin{center} 
\vspace{-0.3cm}
	\begin{tabular}{l  c c c r} 
		Dataset &$N$ &$p$ &$G$ &Disease \\
		\hline
		  Ramaswamy {\it et al.} \cite{ramaswamy2001multiclass}	&190  	&16,063 	&14 	&Cancer \\
		Yeoh {\it et al.}  \cite{yeoh2002classification}	&248  	&12,625		&6 		&Leukemia \\
		 Sun {\it et al.}  \cite{sun2006neuronal}	&180  	&54,613 	&4 		&Glioma \\
		 Nakayama {\it et al.}  \cite{nakayama2007gene}	&105  	&22,283		&10		&Sarcoma \\
		\hline
	\end{tabular}
\end{center}

We compute the results for each dataset over 10 training-test set splits, each with a random choice of training and test set containing 75\% and 25\% of the total $N$ observations, respectively. The average results of classification and gene-selection by CRDA methods are given in Table \ref{table:t3rd} with available comparison results.  
The proposed CRDA based classifiers showcase better classification and feature-selection results for all simulation setups.
Overall, it seems that $\ell_2$ and $\ell_\infty$-norm based CRDA methods are doing better as compared to others.  Moreover, the CRDA based on $\ell_\infty$-norm appears to have best overall performance.  Note that the proposed CRDA classifiers outperform other methods with a significant margin in the case of Ramaswamy {\it et al.} (with 14 groups) and Sun {\it et al.} (of $p=54,613$ genes). 


\section{Discussions and Conclusions} \label{sec:concl}

We proposed a modified version of LDA, called compressive regularized discriminant CRDA, for analysis of data sets in high dimension low sample size situations.  
CRDA was shown to outperform competing methods in most of the cases. It also  had the best detection rate which illustrates that the method can be a useful  tool for accurate selection of (differentially expressed) genes in microarray studies. 

\vfill\pagebreak

\vfill\pagebreak

\end{document}